\documentclass[10pt,twocolumn,aps,prc]{revtex4-2}
\usepackage[T1]{fontenc}
\usepackage{amsmath,bm}
\usepackage{amssymb}
\usepackage{graphicx}
\usepackage{hyperref}
\begin{document}

\title{Hybrid star structure from perturbative QCD}

\author{ Joydev Lahiri$^{1}$ and D. N. Basu$^{1}$}

\affiliation{$^1$Variable Energy Cyclotron Centre, 1/AF Bidhan Nagar, Kolkata 700064, India}

\email[E-mail 1: ]{joy@vecc.gov.in}
\email[E-mail 2: ]{dnb@vecc.gov.in} 

\date{\today }

\begin{abstract}

    The stellar configurations of quark stars are studied using perturbative QCD (pQCD) for the equation of state (EoS). The neutron star structures with equation of state obtained from Brussels-Montreal extended Skyrme interaction are also explored. The deconfinement phase transition from quark to hadron phase in stellar interior for matter under extreme pressure is accomplished by employing the Maxwell construction. The influence of hybrid EoS on the jump in energy density has been investigated. To study hybrid stars the BSk24 hadronic model and pQCD EoS for the quark phase have been used. The properties of hybrid stars in the view of the very recent astrophysical observations have been examined. We find that the gravitational mass might exceed 2.3 $M_\odot$ in some cases, comparable with the observed mass of the pulsar PSR J0952-0607 recently detected.

\vskip 0.2cm
\noindent
{\it Keywords}: Quark stars; Hybrid stars; Equation of state; Phase transition; $\beta$-equilibrated matter.
 
\end{abstract}

\maketitle

\noindent
\section{Introduction}
\label{section1}

    In order to test our understanding on nuclear matter under extreme conditions, a natural laboratory is provided by the study of astrophysical compact objects, especially neutron stars and related pulsars, which are among the most dense objects in the universe. A precise knowledge of the nuclear matter equation of state (EoS) reliable under extreme conditions and a wide range of densities is an essential tool for description of these objects. A succinct exposition of the underlying physics is well described in the textbooks \cite{Teuk04,Glend97,Weber99} which covers our knowledge of the relevant physics till about the 1990s. In order to comprehend such complex structure, one requires the EoS of compact stars in different density regions. For example, the regions of very low density, the sub-nuclear density and from neutron drip density to about nuclear density, can be well described by Feynman-Metropolis-Teller (FMT) \cite{FMT}, Baym-Pethick-Sutherland (BPS) \cite{Ba71} and Baym-Bethe-Pethick (BBP) \cite{Ba71a} EoSs, respectively.

   In the present work NS structure is studied using a composite EoS, i.e. FMT, BPS, BBP and the EoS of the $\beta$-equilibrated dense neutron star matter with progressively increasing densities. The different density regions of a compact star are regulated by different EoSs. The density domain can be largely classified into two distinct regions: a crust which is responsible for $\sim$ 0.5$\%$ of mass and $\sim$ 10$\%$ of the radius of a star. The core accounts for the rest of the mass and radius of the star. Except in the outer few meters, the outer layers consist of a solid crust about a km thick comprising a lattice of bare nuclei immersed in a degenerate electron gas. When one penetrates into the crust deeper, the nuclear species become progressively more neutron rich because of the rising electron Fermi energy.
   
	 Meanwhile, very recent observations of pulsars PSR J0952-0607 \cite{Rom22}, PSR J0740+6620 \cite{Fon21,Leg21,Ril21} and PSR J0348+0432 \cite{An13}, predict their masses $\gtrsim 2 M_\odot$ to a reasonable degree of accuracy. An analysis of the GW170817 event \cite{Rez18} also corroborates similar masses. Motivated by these observations, the structures of quark star (QS), neutron star (NS) and hybrid star (HS) have been studied in detail. For the quark matter perturbative QCD (pQCD) EoS \cite{Kur10,Fra14} has been used. For the hadronic matter, Brussels-Montreal extended Skyrme (BSk) interaction \cite{Kre77,Ge86,Zhu88,Cha09,Gor10,Gor13,Gor15} has been used. Of these BSk family, the BSk24 interaction leads to the highest NS mass. Hence, in all subsequent studies, this interaction is preferred and for the outer crust, FMT+BPS+BBP EoSs have been used. At very high density, the deconfinement phase transition from hadronic matter to quark matter can occur inside NS leading to HS. This phenomenon is characterized by a jump in the energy density at quark-hadron phase transition inside the core which is explored using Maxwell construction. The mass radius relationship of HSs have been studied in the light of recent astrophysical observations.
        
\noindent
\section{Theoretical formalism}
\label{section2}

In this section we introduce the quark and the hadronic matter models separately which will be used in constructing the quark stars, neutron stars and hybrid stars. The equation of state for cold quark matter using perturbative QCD (pQCD) and the extended Skyrme model for the hadronic matter are presented in the following subsections.

\noindent
\subsection{Equation of state for cold quark matter}
   
	The pQCD EoS for cold quark matter \cite{Kur10} in beta equilibrium can be described fairly accurately by the compact fitting function \cite{Fra14} as
	
\begin{eqnarray}
P_{\rm{QCD}}(\mu_B,X) &=& P_{\rm{SB}}(\mu_B) \left( c_1 - \frac{a(X)}{(\mu_B/{\rm GeV}) - b(X)} \right),  \nonumber\\
a(X) &=& d_1 X^{-\nu_1},\quad
b(X) = d_2 X^{-\nu_2}, 
\label{eq:press}
\end{eqnarray}
where $P_{\rm{QCD}}(\mu_B,X)$ is the pressure of the cold quark matter with
\begin{eqnarray}
P_{\rm SB}(\mu_B) = \frac{3}{4\pi^2}(\mu_B/3)^4.
\label{seqn2}
\end{eqnarray}
being the Stefan-Boltzmann limit of the pressure for three non-interacting quark flavors. 

The functions $a(X)$ and $b(X)$ encode the dependence of the result on renormalization scale $\bar{\Lambda}$. The dimensionless parameter $X = 3\bar{\Lambda}/\mu_B$, which can take values from $1$ to $4$. The numerical constants $\{ c_1,d_1,d_2,\nu_1,\nu_2\}$ are given by the best fit \cite{Fra14} values
\begin{eqnarray}
c_1=0.9008 \quad & d_1= 0.5034 &\quad d_2 = 1.452 \nonumber\\
\nu_1 &= 0.3553 \quad \nu_2&= 0.9101.
\label{seqn3}
\end{eqnarray}

    From Eq.~\ref{eq:press} the expression for the energy density $\varepsilon$ can be given by
\begin{equation}
\varepsilon_{\rm{QCD}} =  3P_{\rm{QCD}}+\frac{\mu_B}{{\rm GeV}} P_{\rm{SB}}(\mu_B)   
\frac{a(X)}{\left[ (\mu_B/{\rm GeV}) - b(X)\right]^{2}}. 
\label{eq:edens}
\end{equation}
It is interesting to note that in MIT bag model, $\varepsilon_{\rm{QCD}} - 3P_{\rm{QCD}} = 4B$, where $B$ is  the bag constant. The baryon number density $\rho_B$ can be obtained using the well known thermodynamic relation $\rho_B \mu_B=\varepsilon_{\rm{QCD}} + P_{\rm{QCD}}$.

\noindent
\subsection{The \texorpdfstring{$\beta$}{}-equilibrated cold dense \texorpdfstring{$npe\mu$}{} matter}
    
    The generalized Skyrme interaction for the hadronic matter can be obtained by extending the standard form of the Skyrme interaction \cite{Sk56,Ch97} with additional zero-range density- and momentum-dependent terms~\cite{Kre77,Ge86,Zhu88,Cha09,Gor10,Gor13,Gor15} as  
\begin{eqnarray}
V(\bm{r}_1,\bm{r}_2)&=&t_0(1+x_0P_{\sigma})\delta(\bm{r})  \notag \\
& &+\frac{1}{2}t_1(1+x_1P_{\sigma})[\bm{k}'^2\delta(\bm{r})+\mathrm{c.c.}] \notag \\
& &+t_2(1+x_2P_{\sigma})\bm{k}'\cdot\delta(\bm{r})\bm{k} \notag \\
& &+\frac{1}{6}t_3(1+x_3P_{\sigma})\rho ^{\alpha}\left(\bm{R}\right)\delta(\bm{r})\notag\\
& &+iW_0(\bm{\sigma}_1+\bm{\sigma}_2)\cdot[\bm{k}'\times\delta(\bm{r})\bm{k}] \notag\\
& &+\frac{1}{2}t_4(1+x_4P_{\sigma}) 
\left[\bm{k}'^2\rho^{\beta}\left(\bm{R}\right)\delta(\bm{r})+\mathrm{c.c.} \right]\notag\\
& &+t_5(1+x_5P_{\sigma})\bm{k}'\cdot\rho^{\gamma}\left(\bm{R}\right)\delta(\bm{r})\bm{k} 
\label{Eq:ExSky}
\end{eqnarray}
\noindent
where $\bm{r}=\bm{r}_1-\bm{r}_2$, $\bm{R}=(\bm{r}_1+\bm{r}_2)/2$, $\bm{\sigma}_i$ is the Pauli spin operator, $P_{\sigma}$
is the spin-exchange operator,
$\bm{k}=-i(\bm{ \nabla}_1-\bm{\nabla}_2)/2$ is the relative
momentum operator, and
$\bm{k}^{\prime}$ is the conjugate operator of $\bm{k}$ acting
on the left. In the above equation the first term represents central term, second and third terms represent non-local term, fourth term represents density dependent term and last term represents spin-orbit term. The last two terms are the density dependent generalization of the $t_1$ and $t_2$ terms in the standard Skyrme model. Infinite nuclear matter being spatially homogeneous, the axis can not be defined which leads to the absence of spin-orbit coupling. This implies that $W_0$ term does not contribute in nuclear matter calculations. Thus the generalized Skyrme interaction collectively requires 15 parameters, {\it viz.} $t_0$, $t_1$, $t_2$, $t_3$, $t_4$, $t_5$, $x_0$, $x_1$, $x_2$, $x_3$, $x_4$, $x_5$, $\alpha$, $\beta$ and $\gamma$. The numerical values of these parameters used in the present work are listed in Table-\ref{table1}.Using density functional theory the energy per particle of an asymmetric infinite nuclear matter (ANM) can be expressed as \cite{Pea18}


\begin{multline}
\epsilon(\rho,\eta) = 
\frac{3\hbar^2}{20}\left[\frac{1}{M_\mathrm{n}}(1+\eta)^{5/3} +
\frac{1}{M_\mathrm{p}}(1-\eta)^{5/3}\right]nk_\mathrm{F}^2    \\
 + \frac{1}{8}t_0\Biggl[3 - (1+2x_0)\eta^2\Biggr]n^2  \\
 + \frac{3}{40}t_1\Biggl[(2+x_1)F_{5/3}(\eta) -
\left(\frac{1}{2}+x_1\right)F_{8/3}(\eta) \Biggr]n^2k_\mathrm{F}^2 \\
 + \frac{3}{40}t_2\Biggl[(2+x_2)F_{5/3}(\eta) +
\left(\frac{1}{2}+x_2\right)F_{8/3}(\eta) \Biggr]n^2k_\mathrm{F}^2 \\
 + \frac{1}{48}t_3\Biggl[3-(1+2x_3)\eta^2\Biggr]n^{\alpha+2} \\
 + \frac{3}{40}t_4\Biggl[(2+x_4)F_{5/3}(\eta) - 
\left(\frac{1}{2}+x_4\right)F_{8/3}(\eta) \Biggr]n^{\beta+2}k_\mathrm{F}^2 \\
 + \frac{3}{40}t_5\Biggl[(2+x_5)F_{5/3}(\eta) + 
\left(\frac{1}{2}+x_5\right)F_{8/3}(\eta) \Biggr]n^{\gamma+2}\,k_\mathrm{F}^2 \quad , 
\label{seqn6}
\end{multline}
\begin{equation}
k_\mathrm{F}= \left(\frac{3\pi^2n}{2}\right)^{1/3} \quad  ,
\label{seqn7}
\end{equation}

\begin{equation}
\eta = \frac{\rho_n - \rho_p}{\rho} = 1 - 2Y_p
\label{seqn8}
\end{equation}
and
\begin{equation}
F_m(\eta)= \frac{1}{2}\Biggl[(1+\eta)^m+(1-\eta)^m\Biggr]\quad.
\label{seqn9}
\end{equation}
Here $Y_p=Z/A=\rho_p/\rho$ is the proton fraction, $\rho_p$ and $\rho$ are the proton and baryonic number densities, respectively. The pressure $P(\rho,\eta)$ in ANM may then be expressed as

\begin{multline}
P(\rho,\eta)
=\frac{\hbar^2}{10}\left[\frac{1}{M_\mathrm{n}}(1+\eta)^{5/3}+
\frac{1}{M_\mathrm{p}}(1-\eta)^{5/3}\right]nk_\mathrm{F}^2    \\
+ \frac{1}{8}t_0\Biggl[3 - (1+2x_0)\eta^2\Biggr]n^2  \\
+ \frac{1}{8}t_1\Biggl[(2+x_1)F_{5/3}(\eta) -
\left(\frac{1}{2}+x_1\right)F_{8/3}(\eta) \Biggr]n^2k_\mathrm{F}^2 \\
 + \frac{1}{8}t_2\Biggl[(2+x_2)F_{5/3}(\eta) +
\left(\frac{1}{2}+x_2\right)F_{8/3}(\eta) \Biggr]n^2k_\mathrm{F}^2 \\
+ \frac{(\alpha + 1)}{48}t_3\Biggl[3-(1+2x_3)\eta^2\Biggr]n^{\alpha+2} \\
+ \frac{3\beta + 5}{40}t_4\Biggl[(2+x_4)F_{5/3}(\eta) -
\left(\frac{1}{2}+x_4\right)F_{8/3}(\eta) \Biggr]n^{\beta+2}k_\mathrm{F}^2 \\
+ \frac{3\gamma + 5}{40}t_5\Biggl[(2+x_5)F_{5/3}(\eta) + 
\left(\frac{1}{2}+x_5\right)F_{8/3}(\eta) \Biggr]n^{\gamma+2}\,k_\mathrm{F}^2 \quad.
\label{seqn10}
\end{multline}
      		
\begin{table}[htbp]
\centering
\caption{Values of the parameters of nuclear matter for the Skyrme interactions corresponding to BSk22, BSk24 and BSk26 \cite{Gor13}. Since the parameters $t_2$ and $x_2$ always appear in the form of $t_2x_2$, the value of the latter is provided.}
\begin{tabular}{lccc}
\hline
Parameter  & BSk22& BSk24& BSk26 \\ \hline
$t_0$ [MeV $\mathrm{fm}^{3}$]           &-3978.97  &-3970.29 &-4072.53  \\ 
$t_1$ [MeV $\mathrm{fm}^{5}$]           &404.461   &395.766  &439.536  \\
$t_3$ [MeV $\mathrm{fm}^{3+3\alpha}$]  	  &22704.7   &22648.6  &23369.1  \\
$t_4$ [MeV $\mathrm{fm}^{5+3\beta}$]    &-100.000  &-100.000 &-100.0  \\
$t_5$ [MeV $\mathrm{fm}^{5+3\gamma}$]     &-150.000  &-150.000 &-120.0  \\
$x_0$   &0.16154                 &0.894371  &0.577367  \\
$x_1$   &-0.047986	            &0.0563535 &-0.404961  \\
$t_2x_2$ [MeV $\mathrm{fm}^{5}$]&-1396.13&-1389.61  &-1147.70  \\
$x_3$   &0.514386  &1.05119  &0.624831  \\
$x_4$   	   &2.00000   &2.00000  &-3.00000  \\
$x_5$   	 	 &-11.0000  &-11.0000 &-11.00000  \\
$\alpha$	 	 &1/12  &1/12 &1/12  \\
$\beta$    &1/2   &1/2  &1/6  \\
$\gamma$ &1/12  &1/12 &1/12  \\ 
\hline
\end{tabular}
\label{table1} 
\end{table}

\begin{figure}[t]
\centering
\includegraphics[width=8cm]{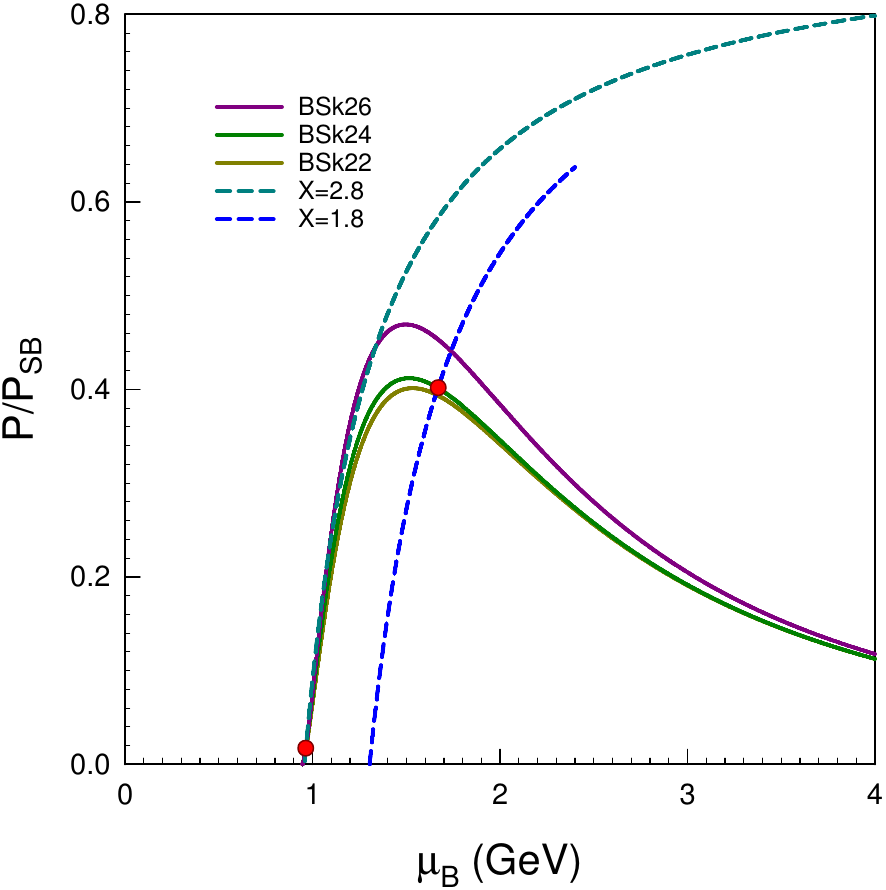}
\caption
{Plots of pressure versus baryonic chemical potential $\mu_B$ of quark matter for different values of parameter $X$. The corresponding plots for hadronic matter are also shown.}
\label{fig1}
\end{figure}

\begin{figure}[t]
\centering
\includegraphics[width=8cm]{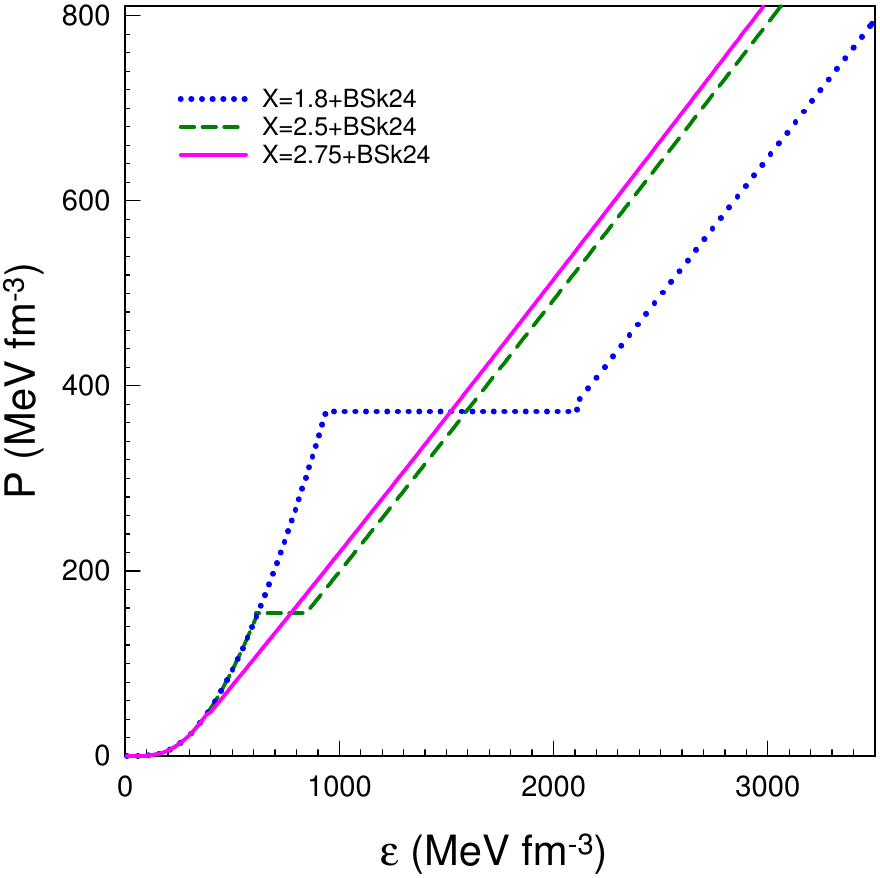}
\caption
{Plots of pressure versus energy density for EoS of hybrid quark–neutron stars. The phase transition for different values of $X$ can be observed by the jumps in energy density except in case of $X=2.75+$BSk24 which is associated with vanishing latent heat.}
\label{fig2}
\end{figure}

\begin{figure}[t]
\centering
\includegraphics[width=8cm]{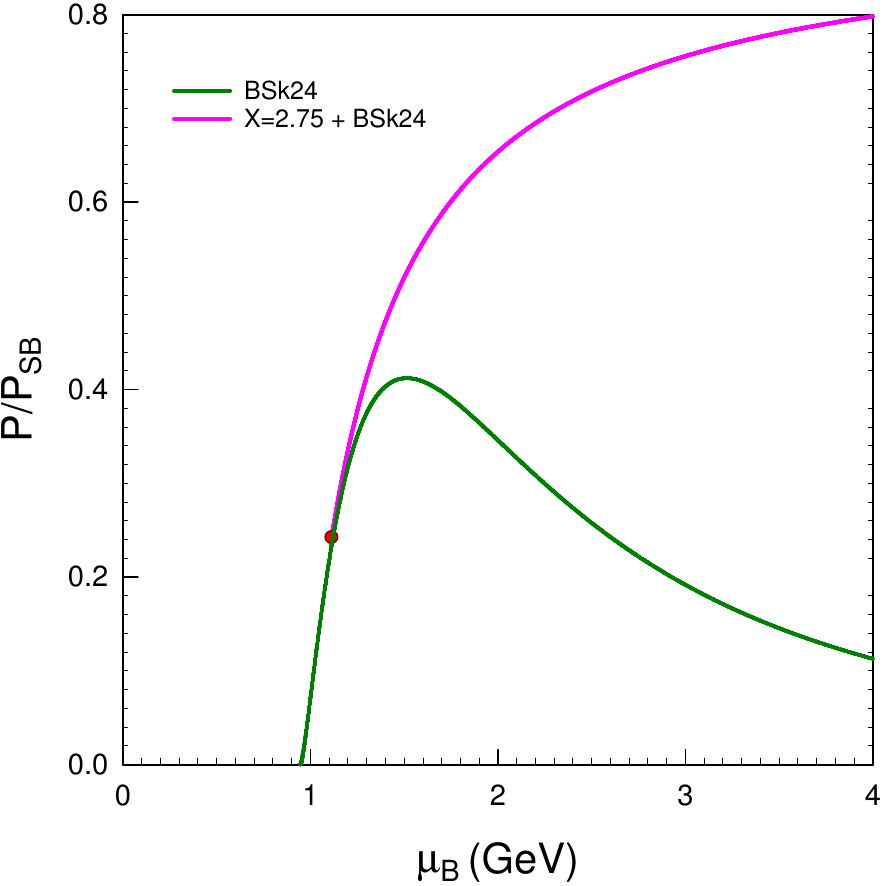}
\caption
{Pressure resulting from the matching of pQCD EoS for $X=2.75$ with that of BSk24 EoS for hadronic matter is shown. The red dot represents the matching point of the two EoSs.}
\label{fig3}
\end{figure}
\noindent  
		
\begin{figure}[t]
\centering
\includegraphics[width=8cm]{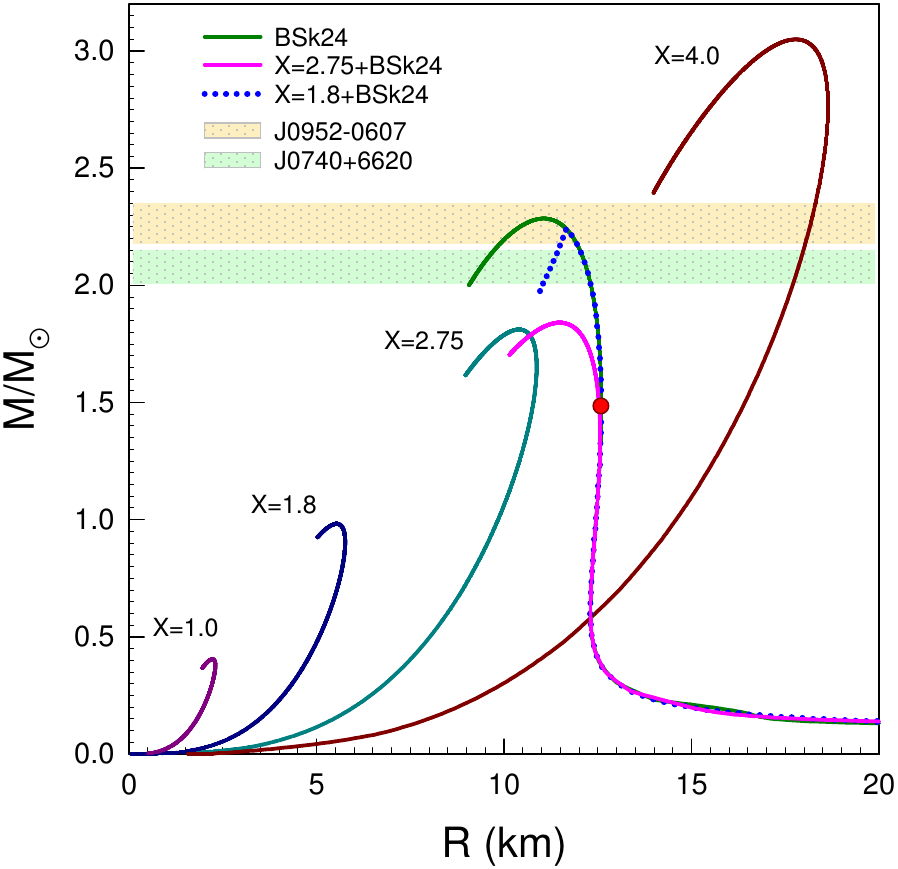}
\caption
{The M-R diagrams for pure quark, hadronic and hybrid quark–neutron stars are shown. The red dot represents the special point. Horizontal bands correspond to mass constraints of PSR J0952-0607 and PSR J0740+6620.}
\label{fig4}
\end{figure}
\noindent 

\noindent
\section{Results and discussion}
\label{section3}

    In general relativity, the structure of a spherically symmetric body of isotropic material which is in static gravitational equilibrium is given by the Tolman-Oppenheimer-Volkoff (TOV) equation \cite{TOV39a,TOV39b} 

\begin{eqnarray}
\frac{dP(r)}{dr} = -\frac{G}{c^4}\frac{[\varepsilon(r)+P(r)][m(r)c^2+4\pi r^3P(r)]}{r^2[1-\frac{2Gm(r)}{rc^2}]}, \nonumber \\ 
\varepsilon(r)=(\epsilon + m_b c^2)\rho(r),~m(r)c^2=\int_0^r \varepsilon(r') d^3r', 
\label{seqn11}
\end{eqnarray}
\noindent 
where $\varepsilon(r)$ and $P(r)$ are, respectively, the energy density and the pressure at a radial distance $r$ from the center of the star and $m(r)$ is the stellar mass contained within radius $r$. The numerical solution of TOV equation for masses and radii has been obtained using Runge-Kutta method. The EoS provides $\varepsilon(\rho)$ and $P(\rho)$ as the inputs for the calculations. The boundary condition $P(r)=0$ at the surface $R$ of the star determines its size and integration up to $R$ given by $M=m(R)$ \cite{Um97} provides the its total mass $M$. The numerical solution of TOV equation, being an initial value problem, requires a single integration constant ${\it viz.}$ the pressure $P_c$ at the center $r=0$ of the star calculated at a given central density $\rho_c$. It is important to mention here that the solution of TOV equation determines masses of static stars which are very close to slowly rotating NSs \cite{Ch10,Mi12,Ba14}.   

\noindent
\subsection{Deconfinement phase transition: from Hadronic to Quark matter}

    In the present work, the quark-hadron deconfinement phase transition is obtained using Maxwell construction \cite{Mar07}, following which the transition from hadronic to quark phase occurs with sharp jump in energy density at a point where the pressure and baryonic chemical potential of the individual charge neutral phases are equal \cite{Sch16,Len12,Bha10,Gom19,Han19,Fer20} implying 
		
\begin{eqnarray}
P_{\rm{QCD}}(\mu_B^Q,X)&=& P(\rho,\eta) \nonumber \\
\mu_B^Q&=&\mu_B^H. 
\label{seqn12}
\end{eqnarray}
\noindent			
			
In Fig.~\ref{fig1} the pressure versus baryonic chemical potential $\mu_B$ of quark matter for different values of parameter $X$ have been plotted. The corresponding plots for hadronic matter are also shown. The red dots satisfy the conditions of Eq.~\ref{seqn12} for hadronic EoS BSk24 and quark EoSs corresponding to $X=$1.8 and 2.8.  
 
In Fig.~\ref{fig2} plots of pressure versus energy density for EoS of hybrid quark–neutron stars are shown. The phase transitions for $X=1.8+$BSk24 and $X=2.5+$BSk24 can be observed by the jumps in energy density. In case of $X=2.75+$BSk24 this scenario is not observed implying that it is associated with continuous pressure and vanishing latent heat.

In Fig.~\ref{fig3} pressure resulting from the matching of pQCD EoS for $X=2.75$ with that of BSk24 EoS for hadronic matter is shown. The red dot represents the matching point of the two EoSs. It is interesting to note that the behaviors of the BSk24 EoS and quark matter ($X=2.75$) EoS are markedly similar in the region near the matching point.

\subsection{The mass-radius relationship for quark stars, neutron stars and hybrid stars}

		The mass-radius (M-R) relationship has been obtained for quark stars by solving Tolman-Oppenheimer-Volkoff (TOV) equation using quark matter EoS for different values of the dimensionless parameter $X$. For the outer crust region in case of NS and HS, the EoSs FMT \cite{FMT}, BPS \cite{Ba71} and BBP \cite{Ba71a} up to the number density of 0.0582 fm$^{-3}$ have been used. For the inner crust and beyond the $\beta$-equilibrated NS matter (hybrid EoS) have been used to study the structural properties of the NS (HS). The hybrid EoS used for HS have been computed for a few different values of $X$.
		
		In Fig.~\ref{fig4}, the M-R relationship for quark stars have been plotted for $X=$1.0, 1.8, 2.75 and 4.0. The maximum masses and corresponding radii are 0.41 $M_\odot$ and 2.21 km, 0.98 $M_\odot$ and 5.53 km, 1.81 $M_\odot$ and 10.40 km, 3.05 $M_\odot$ and 17.78 km, respectively. As may be seen in the figure that the M-R curves for quark stars are distinctly different from those of NSs and HSs.
    
    The M-R relationship for slowly rotating NSs for EoSs obtained with BSk24 is also plotted in Fig.~\ref{fig4} for different values of the parameter $X$. The maximum NS mass for the EoS obtained using the BSk24 Skyrme set is 2.28 $M_\odot$ with a corresponding radius of 11.05 kms where $M_\odot$ is the solar mass. The horizontal bands in this figure correspond to mass constraints of PSR J0740+6620 and PSR J0952-0607 \cite{Fon21,Leg21,Ril21,Rom22}. 
		
		The M-R plots for HSs show almost similar pattern as those of NSs with maximum masses being smaller. The maximum masses and corresponding radii for BSk24 with quark cores for $X=$2.75 and $X=$1.8 are 1.84 $M_\odot$ and 11.48 km, 2.24 $M_\odot$ and 11.66 km, respectively. 
		
    It is important to mention here that the observations of the two-solar-mass binary millisecond pulsar J1614-2230 by Demorest et al.~\cite{De10} suggest that the masses lie within 1.97 $\pm$ 0.04 $M_\odot$, effectively excluding most of the hyperon or boson condensate equations of state. The radio timing measurements of the pulsar PSR J0348+0432 and its white dwarf companion have confirmed the mass of the pulsar to be in the range 2.01 $\pm$ 0.04 $M_\odot$ \cite{An13}. Very recently, the studies for PSR J0740+6620 \cite{Fon21} and for PSR J0952-0607 \cite{Rom22} find masses of 2.08 $\pm$ 0.07 $M_\odot$ and 2.35 $\pm$ 0.17 $M_\odot$, respectively. Some recent works \cite{Leg21,Ril21} constrain the equatorial radius and mass of PSR J0740+6620 to be 12.39$^{+1.30}_{-0.98}$ km and 2.072$^{+0.067}_{-0.066}$ $M_\odot$ respectively. From the recent observation of the gravitational wave event GW170817, the limits for the maximum mass has been estimated as ${2.01}_{-0.04}^{+0.04}\leqslant {M}_{\mathrm{TOV}}/{M}_{\odot }\lesssim {2.16}_{-0.15}^{+0.17}$ \cite{Rez18}.  The maximum NS masses attained for BSk22, BSk24 and BSk26 are, respectively, 2.27 $M_\odot$, 2.28 $M_\odot$ and 2.18 $M_\odot$ which are in accordance with recent observations of massive NSs. 			

\noindent
\section{ Summary and conclusion }
\label{section4}

In summary, the structure of quark stars, neutron stars and hybrid stars have been explored. Using Maxwell construction the quark to hadron deconfinement phase transition in neutron star cores and the formation of hybrid stars have been investigated. The result for neutron star without quark core is in excellent agreement with recent astrophysical observations comparable with the observed mass 2.3 $M_\odot$ of the pulsar PSR J0952-0607. The compact star matter can undergo deconfinement transition to quark matter, reducing its mass considerably. The calculated hybrid star properties also agree with recent astrophysical constraints on the M-R relationship obtained from PSR J0952-0607, PSR J0740+6620, PSR J0348+0432 and the GW170817 data analysis.
   
\begin{acknowledgments}

    One of the authors (DNB) acknowledges support from Science and Engineering Research Board, Department of Science and Technology, Government of India, through Grant No. CRG/2021/007333.

\end{acknowledgments}


\noindent
\begin{thebibliography}{99}

\bibitem{Teuk04} Stuart L. Shapiro, and Saul A. Teukolsky, Black Holes, White Dwarfs, and Neutron Stars: The Physics of Compact Objects, WILEY‐VCH Verlag GmbH. (2004).

\bibitem{Glend97} Norman K. Glendenning, Compact Stars, Springer-Verlag New York, Inc. (1997). 

\bibitem{Weber99} Fridolin Weber, Pulsars as Astrophysical Laboratories for Nuclear and Particle Physics, CRC Press. (1999).

\bibitem{FMT} R. P. Feynman, N. Metropolis and E. Teller, Phys. Rev. {\bf 75}, 1561 (1949).

\bibitem{Ba71} G. Baym, C. Pethick and P. Sutherland, Astrophys. J. {\bf 170}, 299 (1971).

\bibitem{Ba71a} G. Baym, H. A. Bethe, and C. J. Pethick, Nucl. Phys. {\bf A 175}, 225 (1971).

\bibitem{Rom22} Roger W. Romani et al., Astrophys. J. Lett. {\bf 934}, L17 (2022).

\bibitem{Fon21} E. Fonseca et al., Astrophys. J. Lett. {\bf 915}, L12 (2021).

\bibitem{Leg21} Isaac Legred, Katerina Chatziioannou, Reed Essick, Sophia Han and Philippe Landrya, Phys. Rev. {\bf D 104}, 063003 (2021).

\bibitem{Ril21} Thomas E. Riley et al., Astrophys. J. Lett. {\bf 918}, L27 (2021).

\bibitem{An13} J. Antoniadis et al., Science {\bf 340}, 6131 (2013).

\bibitem{Rez18} Luciano Rezzolla, Elias R. Most and Lukas R. Weih, Astrophys. J. Lett. {\bf 852}, L25 (2018).

\bibitem{Kur10} A. Kurkela, P. Romatschke, and A. Vuorinen, Phys. Rev. D \textbf{81}, 105021 (2010).

\bibitem{Fra14} E. S. Fraga, A. Kurkela, and A. Vuorinen, Astrophys. J. {\bf 781}, L25 (2014).

\bibitem{Kre77} S. Krewald, V. Klemt, J. Speth, and A. Faessler, Nucl. Phys. \textbf{A281}, 166 (1977).

\bibitem{Ge86} L.X. Ge, Y. Z . Zhuo, and W. Norenberg, Nucl. Phys. \textbf{A459}, 77 (1986).

\bibitem{Zhu88} Y.Z. Zhuo, Y.L. Han, and X.Z. Wu, Prog. Theor. Phys. \textbf{79}, 110 (1988).

\bibitem{Cha09} N. Chamel, S. Goriely, and J.M. Pearson, Phys. Rev. C \textbf{80}, 065804 (2009).

\bibitem{Gor10} S. Goriely, N. Chamel, and J.M. Pearson, Phys. Rev. C \textbf{82}, 035804 (2010).

\bibitem{Gor13} S. Goriely, N. Chamel, and J.M. Pearson, Phys. Rev. C \textbf{88}, 024308 (2013).

\bibitem{Gor15} S. Goriely, Nucl. Phys. \textbf{A933}, 68 (2015).

\bibitem{Sk56} T. H. R. Skyrme, Phil. Mag. {\bf l}, 1043 (1956); {\it ibid} Nucl. Phys. {\bf 9}, 615 (1959).

\bibitem{Ch97} E. Chabanat, E Bonche, E Haensel J. Meyer and R. Schaeffer, Nucl. Phys. {\bf A 627}, 710 (1997).

\bibitem{Pea18} J. M. Pearson, N. Chamel, A. Y. Potekhin, A. F. Fantina, C. Ducoin, A. K. Dutta and S. Goriely, Mon. Not. R. Astron. Soc. {\bf 481}, 2994 (2018).

\bibitem{TOV39a} R. C. Tolman, Phys. Rev. {\bf 55}, 364 (1939).

\bibitem{TOV39b} J. R. Oppenheimer and G. M. Volkoff, Phys. Rev. {\bf 55}, 374 (1939).

\bibitem{Um97} V. S. Uma Maheswari, D. N. Basu, J. N. De and S. K. Samaddar, Nucl. Phys. {\bf A 615}, 516 (1997).

\bibitem{Ch10} P. R. Chowdhury, A. Bhattacharyya and D. N. Basu, Phys. Rev. {\bf C 81}, 062801(R) (2010).

\bibitem{Mi12} Abhishek Mishra, P. R. Chowdhury and  D. N. Basu, Astropart. Phys. {\bf 36}, 42 (2012).

\bibitem{Ba14} D. N. Basu, Partha Roy Chowdhury and Abhishek Mishra, Eur. Phys. J. Plus {\bf 129}, 62 (2014).

\bibitem{Mar07} T. Maruyama, S. Chiba, H.-J. Schulze and T. Tatsumi, Phys. Lett. {\bf B 659}, 192 (2008); {\it ibid} Phys. Rev. {\bf D 76}, 123015 (2007).

\bibitem{Sch16} S. Schramm, V. Dexheimer and R. Negreiros, Eur. Phys. J. {\bf A 52}, 14 (2016).

\bibitem{Len12} C. H. Lenzi and G. Lugones, Astrophys. J. {\bf 759}, 57 (2012).

\bibitem{Bha10} A. Bhattacharya, I. N. Mishustin and W. Greiner, J. Phys. {\bf G 37}, 025201 (2010).

\bibitem{Gom19} R. O. Gomes, P. Char and S. Schramm, Astrophys. J. {\bf 877}, 139 (2019).

\bibitem{Han19} S. Han, M. A. A. Mamun, S. Lalit, C. Constantinou and M. Prakash, Phys. Rev. {\bf D 100}, 103022 (2019).

\bibitem{Fer20} M. Ferreira, R. C. Pereira, and C. Providencia, Phys. Rev. {\bf D 101}, 123030 (2020).

\bibitem{De10} P. B. Demorest, T. Pennucci, S. M. Ransom, M. S. E. Roberts, J. W. T. Hessels, Nature {\bf 467}, 1081 (2010).

\end{thebibliography}
\end{document}